\include{graphicsx} 
\documentclass[12pt,preprint]{aastex}

\newcommand{\kms}{\mbox{ km s$^{-1}$}}

\slugcomment{}
\shorttitle{MagAO H$\alpha$ Images of the Binary Proplyd LV 1}
\shortauthors{Wu et al.}

\begin{document}

\title{High Resolution H alpha Images of the Binary Low-mass Proplyd LV 1 with the Magellan AO System\altaffilmark{*}} 

\author{Y.-L. Wu$^1$, L. M. Close$^1$, 
 J. R. Males$^{1,4}$,
 K. Follette$^1$,
 K. Morzinski$^{1,4}$, 
 D. Kopon$^{1,2}$, 
 T. J. Rodigas$^1$, 
 P. Hinz$^1$,
 A. Puglisi$^3$, 
 S. Esposito$^3$,
 E. Pinna$^3$,
 A. Riccardi$^3$,
 M. Xompero$^3$, \and
 R. Briguglio$^3$
}

\affil{$^1$Steward Observatory, University of Arizona, Tucson, AZ 85721, USA;}
\email{yalinwu@email.arizona.edu}
\affil{$^2$Max Planck Institute for Astronomy, K$\ddot{o}$enigstuhl 17, D-69117 Heidelberg, Germany}
\affil{$^3$INAF -- Osservatorio Astrofisico di Arcetri, Largo E. Fermi 5, I-50125, Firenze, Italy}
\altaffiltext{*}{This paper includes data gathered with the 6.5 m Magellan Telescopes located at the Las Campanas Observatory, Chile.}
\altaffiltext{4}{NASA Sagan Fellow}

\begin{abstract} 
We utilize the new Magellan adaptive optics system (MagAO) to image the binary proplyd LV 1 in the Orion Trapezium at H$\alpha$. This is among the first AO results in visible wavelengths. The H$\alpha$ image clearly shows the ionization fronts, the interproplyd shell, and the cometary tails. Our astrometric measurements find no significant relative motion between components over $\sim$18 yr, implying that LV 1 is a low-mass system. We also analyze Large Binocular Telescope AO observations, and find a point source which may be the embedded protostar's photosphere in the continuum. Converting the $H$ magnitudes to mass, we show that the LV 1 binary may consist of one very-low-mass star with a likely brown dwarf secondary, or even plausibly a double brown dwarf. Finally, the magnetopause of the minor proplyd is estimated to have a radius of 110 AU, consistent with the location of the bow shock seen in H$\alpha$.
\end{abstract}


\section{INTRODUCTION}

The Orion Nebula (M42, NGC1976), being the nearest \citep[$414\pm7$ pc;][]{M07} massive star-forming region, is ideal for investigating the physics of young stellar objects (YSOs) and their interactions with ambient interstellar medium (ISM). \cite{LV79} first discovered several gaseous knots (LV 1--6) adjacent to the Trapezium cluster. These knots are bright in optical emission lines and believed to be protoplanetary disks \citep[``proplyds";][]{OWH93} embedded in primeval material. The UV radiation from the primary Trapezium ionizing star, $\theta^1$ Ori C ($V = 5.13$ mag), constantly illuminates and photoevaporates these circumstellar disks, driving $\sim$3\kms photoevaporated flows \citep{JHB98}. The flows accelerate to few tens of $\kms$ as they pass through the ionization fronts located at several disk radii. The interaction between the outflows and the stellar wind from $\theta^1$ Ori C has created various features, including stationary bow shocks and cometary tails pointing away from the ionizing star. So far in the Orion Nebula more than 200 proplyds, either externally illuminated by ionizing photons from young massive stars or seen in silhouettes against the nebular background, have been discovered by various {\it Hubble Space Telescope (HST)} surveys \citep{B98,B00,R08}.

Of these proplyds the LV 1 is unique as a tight binary system \citep{F93}. Two proplyds separated by 0\farcs4 (projected on the sky) comprise this system, with the secondary (168--326 NW) being physically closer to $\theta^1$ Ori C, and the primary (168--326 SE) located at the southeast. LV 1 has the least massive proplyds in the submillimeter survey carried out by \cite{MW10}, with the total disk mass $<$2 $\times$ $10^{-4} M_{\sun}$. Both proplyds exhibit classic crescent ionization fronts and cometary tails; however, there is also a surprisingly thin straight ``shell'' in-between the components. The shell is prominent in emission lines, e.g., [\ion{O}{3}] and H$\alpha$ \citep{B98}. It was also imaged in 5 GHz radio continuum by \cite{G02}. Simulations suggested that the collision of photoevaporated flows could form such a flat interproplyd shell \citep{H02,V10}, making LV 1 the first evidence of proplyd interaction in the Orion Nebula. This is an important example of a short-lived, but critical, phase of binary evolution in OB clusters.

In this paper, we present high spatial resolution H$\alpha$ observations of LV 1 with the Magellan adaptive optics system (MagAO). MagAO is a new adaptive optics system used with the 6.5 m Clay Telescope. Utilizing a 585-actuator adaptive secondary mirror to correct phase distortions, MagAO is the world's first AO system which demonstrates reasonable performance in visible wavelengths, with Strehl ratio $\sim$30\% and resolution $\sim$20 mas on bright stars $V < 9$ mag \citep[for details of MagAO's design and performance refer to][and references therein]
{C12a, C13}. We also include published near-infrared (near-IR) [\ion{Fe}{2}] and Br$\gamma$ AO observations with the 8.4 m Large Binocular Telescope \citep[LBT;][]{C12b}.

\section{OBSERVATIONS AND REDUCTIONS}

\begin{figure}
\includegraphics[angle=0,width=\columnwidth]{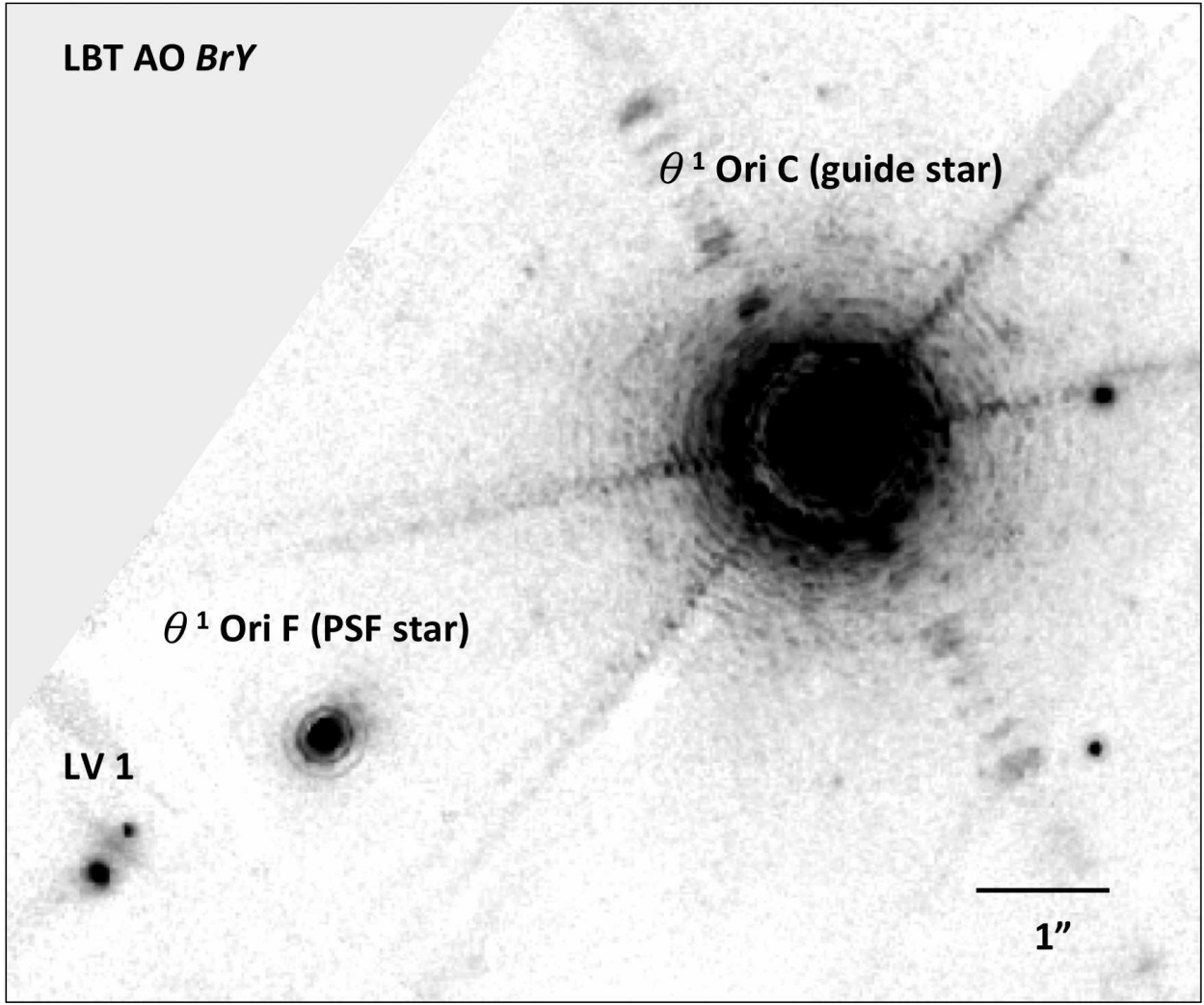}\caption{
LBT AO Br$\gamma$ (2.16 $\mu$m) image showing the positions of $\theta^1$ Ori C, $\theta^1$ Ori F, and LV 1. North is up and east is to the left. Note the direct line between the guide star and the PSF star $\theta^1$ Ori F. Ignore the faint symmetric speckles around $\theta^1$ Ori C, as they are very faint PSF artifacts.
}
\label{fig1}
\end{figure}

MagAO observations of LV 1 at 6563 \AA \, H$\alpha$ and 6730 \AA \, [\ion{S}{2}] were made on 2012 December 3 (UT) during the first MagAO commissioning run. We locked the AO system on $\theta^1$ Ori C at 250 modes and 990 Hz bandwidth.\footnote[5]{Now MagAO fully controls 378 modes, but during the first commissioning run (this data set) only 250 modes were corrected.} The guide star is $\sim$6\farcs3 from the proplyd (Figure \ref{fig1}), but we placed it outside of the $8\arcsec$ field of view of the VisAO science camera \citep{K12, M12, C12a} to prevent saturation over the entire image. As a result of anisoplanatism, the data suffered a mild oval distortion toward $\theta^1$ Ori C with FWHM ranging from 0\farcs05 to 0\farcs06, comparable to {\it HST} performance.

We employed the spectral differential imaging (SDI) mode of the VisAO camera to simultaneously observe the emission line and the continuum (point-spread function, PSF). The SDI mode works by separating the incoming light into two orthogonal linearly polarized beams with a Wollaston prism. The beams then pass two narrowband filters (very near the {\it f}/52 VisAO focus), one for the emission line and the other for the continuum, before arriving the camera \citep[for more on the SDI technique, see][]{C05,C12a,B07}. We obtained $60\times 10$ s H$\alpha$ exposures and $24\times 10$ s at [\ion{S}{2}]. We did not detect LV 1 in the H$\alpha$ continuum nor in [\ion{S}{2}], so these images are not presented. Data reduction is carried out using standard IRAF tasks. The data were first dark-subtracted and cross-correlated, but not flat-field-corrected due to a lack of H$\alpha$ flat frames. However, the VisAO science grade E2V CCD is quite flat to $\la$1\%.  Then we rotated the images counterclockwise by ROTOFF+89.11 deg, where ROTOFF is the standard parallactic angle keyword in the VisAO header, so that north is up and east is left to $\pm 0\fdg3$. After median-combining the data, we selected $\theta^1$ Ori F ($V = 10.12$ mag; $\sim$2$\arcsec$ from the proplyds) as the PSF. Fortuitously, $\theta^1$ Ori F is aligned almost exactly at the same position angle (P.A.) as LV 1 with respect to $\theta^1$ Ori C, so the PSF shows the same isoplanatic elongation and orientation (see Figure \ref{fig1}). Finally we deconvolved the reduced images using the IRAF stsdas $lucy$ task.

The LBT AO reduced images of [\ion{Fe}{2}] (1.644 $\mu$m; FWHM = 0\farcs042) and Br$\gamma$ (2.16 $\mu$m; FWHM = 0\farcs055) from \cite{C12b} were also examined here. We also deconvolved these images with $lucy$.

\section{DISCUSSION}
\subsection{Optical and Near-infrared Images}
LV 1 is unique in its various features sculptured by the powerful stellar wind of $\theta^1$ Ori C and the mutual interactions between two proplyds. Our deconvolved H$\alpha$ image of LV 1 is presented in Figure \ref{fig2}. Features like the bright ionization fronts, the interproplyd shell, and the cometary tails are clearly seen. It is remarkable that we can perform a 6\farcs3 off-axis guiding at H$\alpha$ from the ground and still have excellent images. This also demonstrates the power and potential of MagAO in the visible regime even for off-axis science.

We examine the orbital motion of LV 1 by comparing the MagAO H$\alpha$ \citep[$0\farcs00798\pm0\farcs00002\mbox{ pixel}^{-1}$;][]{C13} with the {\it HST} H$\alpha$ observations of 1995 March 21 \citep{B98}. The MagAO platescale was calibrated by comparing the positions of four Trapezium stars from VisAO images with unsaturated {\it HST} Advanced Camera for Surveys astrometry \citep{R07, C12b}. Since the photospheres are invisible in H$\alpha$, we look for any offset in the ionization fronts. The {\it HST} separation and P.A. were measured to $376\pm5$ mas and $56\fdg5\pm0\fdg5$, while for MagAO we have $371\pm1.3$ mas and $55\fdg8\pm0\fdg3$. The uncertainties include difficulty in fitting the crescent shape, centroid error, platescale error, and geometric distortion. Hence we see over $\sim$18 yr a motion of just $5\pm5.2$ mas and $0\fdg7\pm0\fdg6$. These low values are consistent with no significant motion and quite reasonable if these objects are of low mass since one would expect about 8 mas (maximum velocity on sky with an edge-on circular orbit and masses $0.07 M_{\sun}$) or 1\fdg2 (face-on orbit) relative motion. If, for example, they were each solar mass objects we would expect about 29 mas or 4\fdg5 of motion that we would have detected. Therefore, the lack of any significant offset observed is further evidence of the likely low mass of the system.

LV 1 is almost absent in our optical continuum images, indicating that it is obscured by dust. Recent SOFIA mid-IR observations found evidence that LV 1 shows a NE--SW elongation in accordance with the interproplyd shell \citep{S12}. This implies that dust may be entrained by the photoevaporated flow such that the mid-IR is mainly from the superheated disk atmospheres and the interproplyd shell, not from photospheres of LV 1 itself. However, as the total disk mass is just $<$$2 \times 10^{-4} M_{\sun}$, the extinction may not be very large.

Figure \ref{fig3} shows the deconvolved [\ion{Fe}{2}] (1.644 $\mu$m) and Br$\gamma$ (2.16 $\mu$m) images. The ionization front of 168--326 SE and the interproplyd shell are prominent in Br$\gamma$, but the shell is extremely faint in [\ion{Fe}{2}], as we expect since this should mainly trace continuum emission. On the other hand, 168--326 SE in [\ion{Fe}{2}] looks more like a point source with an offset in position located behind the ionization front seen in Br$\gamma$. Compared with optical observations, [\ion{Fe}{2}] suffers less dust extinction and may possibly detect the photosphere of the embedded protostar itself. Previous efforts to estimate the binary mass in the near-IR have not had our spatial resolution. For instance, \cite{MS94} estimated the total embedded stellar mass at $\sim$0.1 $M_{\sun}$ based on $K'$ measurements. Their value may be an overestimate of the mass as a significant amount of the $K'$ flux seems to come from the bright ionization fronts of the proplyds, judging from our Br$\gamma$ image. 

Therefore, we estimate the mass of protostars by comparing their $H$ magnitudes with stellar evolutionary models. The $H$ magnitudes are $13.15\pm0.21$ and $13.33\pm0.21$ for the major and the minor proplyd, respectively \citep{P98}. Estimating that half the radiation originates from the ionization front as seen in Figure \ref{fig3}, we obtain absolute $H$ magnitudes of $5.82\pm0.30$ and $6.00\pm0.30$ ($D=414\pm7$ pc). This corresponds to $0.073^{+0.012}_{-0.011} \, M_{\sun}$ and $0.067^{+0.011}_{-0.009} \, M_{\sun}$ from the 1 Myr DUSTY tracks of \cite{B02}, given that the age of the Orion Nebula Cluster is $\sim$1 Myr \citep{H97}. We are likely to underestimate the true mass as we do not consider $H$ extinction. However, it should be mild due to small disk mass. Our mass estimate is also consistent with the fact that we see no orbital motion in the H$\alpha$ image. Thus, LV 1 may comprise one very-low-mass star and one brown dwarf, or even two brown dwarfs. We list the properties of LV 1 in Table \ref{tbl-1}.

\subsection{The Location of Magnetopause}

Magnetic field is found to be energetically important in star formation and evolution \citep{C99,D05,GRM06}. It provides extra support to slow down gravitational collapse of molecular clouds, influencing disk structure and planet formation. It also plays a major role in angular momentum transfer via jets and outflows. Therefore, the morphology of YSOs and the embedded magnetic field are deeply linked in a way that the observed features may help trace and constrain the field profile. In this section, we estimate the location of the magnetopause of 168--326 NW based on empirical relations, and show that the result well matches the observations.

The standoff distance of magnetopause can be found by equating the ram pressure of stellar wind and the magnetic pressure, assuming a pure dipole field \citep{CF31}. However, plasma in the ISM will also introduce its own magnetic field, making the dipole assumption invalid. Therefore, we have to account for the contribution from the disk. We first notice that the field strength $B$ correlates with the number density $n$. Zeeman measurements have shown that $B \propto n^{0.47}$ over a wide range of number densities, from clumps of $10^3 \mbox{ cm}^{-3} $ to dense cores of $10^9 \mbox{ cm}^{-3}$ \citep{TH86,C99,V08}.

The radial dependence of surface densities $\Sigma$ is ambiguous due to incomplete understanding of viscosity. Different theories indicate scaling relations ranging from $r^{-0.5}$ to $r^{-2}$ \citep{H98,S07,VB07}. For simplicity, we consider a steady-state rotating disk in vertical hydrostatic equilibrium. \cite{P81} showed that in this scenario the surface density and the scale height $H$ have the relations: $\Sigma \propto T^{-1}_c r^{-3/2}$ and $H \propto \sqrt{T_c r^3}$, where $T_c$ is the midplane gas temperature. Assuming an optically thin surface with an optically thick midplane, \cite{C01} found that $T_c$ scales as $r^{-3/7}$. Therefore, $\Sigma \propto r^{-15/14}$, $H \propto r^{9/7}$, and $n \propto \Sigma/H \propto r^{-2.36}$. These scaling relations are consistent with observations in the Ophiuchus star-forming region, where thermal continuum measurements of dust grains show that $T \propto r^{-(0.5-0.6)}$ and $\Sigma \propto r^{-0.9}$ within a characteristic radius $r_c \sim$ 20--200 AU \citep{A09}. However, \cite{A09} also showed that in regions beyond $r_c$ the surface density often has an exponential taper, $\Sigma \propto \exp(-r/r_c)$. We further notice that the electron number density outside of the ionization front decreases as $r^{-2}$ \citep{B98}. Combining these relations, we find that magnetic field scales as $r^{-1.11}$ for the inner disk, $\exp(-0.47r/r_c)r^{-0.6}$ for the outer disk, and $r^{-0.94}$ beyond the ionization front. 

\begin{figure}
\includegraphics[angle=0,width=\columnwidth]{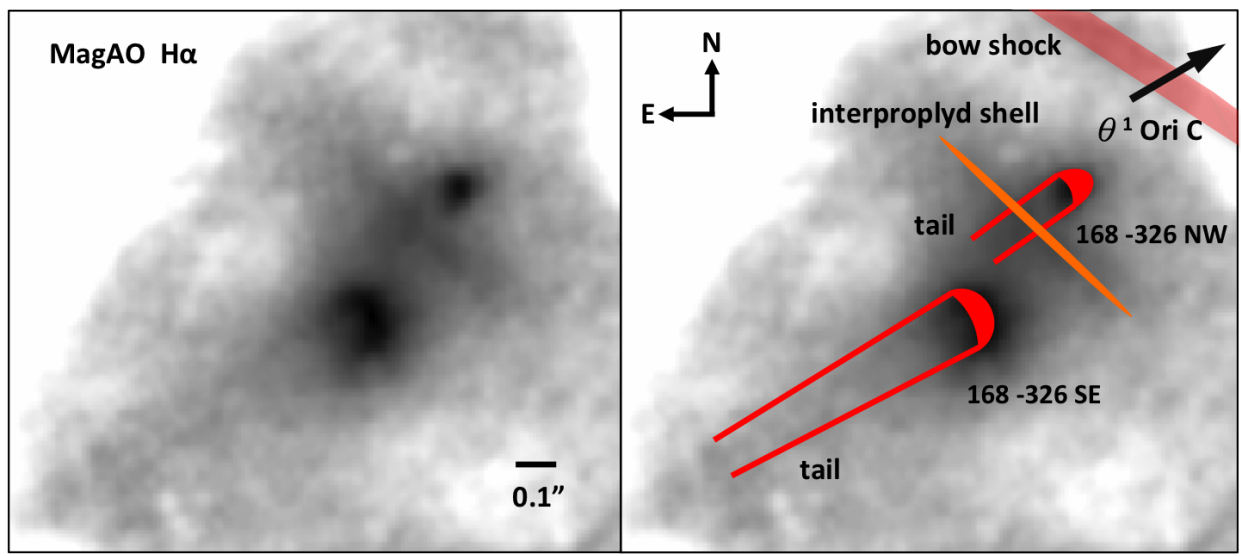}\caption{
Left: the LV 1 as imaged with the MagAO in H$\alpha$ (deconvolved). Right: schematic of LV 1. North is up and east is left in both images.
}
\label{fig2}
\end{figure}

\begin{figure}
\includegraphics[angle=0,width=\columnwidth]{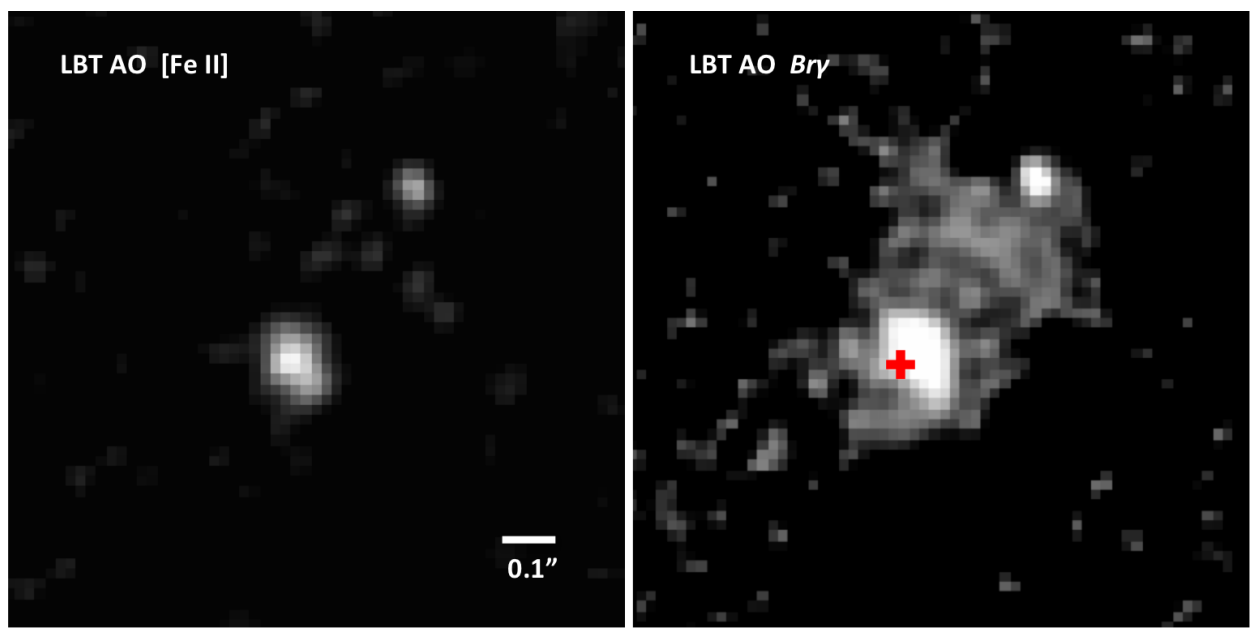}\caption{
Left: deconvolved LBT AO continuum [\ion{Fe}{2}] (1.644 $\mu$m) of LV 1. Right: LBT AO Br$\gamma$ (2.16 $\mu$m) image. The red cross marks the position of a point source seen in the left panel. PSF fitting suggests that about 50\% of the $H$ flux is from the photospheres of the stars, and the rest is extended emission. North is up and east is left.
}
\label{fig3}
\end{figure}

To determine the proportionality constant, we assume $r_c = 5$ AU for 168--326 NW as it is a tiny proplyd. We also assume that its ionization front is at 20 AU, judging from Figure \ref{fig2}. In addition, all brown dwarfs have a size similar to that of Jupiter, and a $0.07 M_{\sun}$ young brown dwarf can have a strong magnetic field $\sim$2 kG on its surface \citep{RC10}. Therefore, we have this empirical formula:  $B(r) = 7.6 \times 10^{-3}(20/r)^{0.94}$ G for $r > 20$ AU.

We finally equate the magnetic pressure with the ram pressure of stellar wind: $\rho v^2 = \dot{M}v/4\pi d^2 = 2B^2/\mu_0$, where $\rho$ and $v$ are the mass density and the speed of stellar wind, $\dot{M}$ is the mass loss rate, $d$ is the distance from the ionizing star, and $\mu_0$ is the vacuum permeability. The mass loss rate of $\theta^1$ Ori C is $\sim$$4 \times 10^{-7} M_{\sun}$ \citep{HP89}, and the wind speed is $\sim$2500$\kms$ \citep{S96}. Substituting these numbers into the above equation, we obtain $r \sim 110$ AU, which is in reasonable agreement with the location of the bow shock ($\sim$150 AU) in Figure \ref{fig2}. This calculation, though oversimplified, demonstrates the possibility in applying the observed features of proplyds to constraining the magnetic field profile.

\section{Summary}
We utilize the newly commissioned MagAO system to image the binary proplyd LV 1 in H$\alpha$. MagAO utilizes an adaptive secondary mirror to achieve a moderate Strehl ratio in the visible. The SDI mode of the VisAO camera enables one to simultaneously observe the line emission and its adjacent continuum. 

We used $\theta^1$ Ori F as the PSF to perform deconvolution. Various structures in LV 1 are identified, including the bow shock, the interproplyd shell, the ionization fronts, and the cometary tails. We find no significant orbital motion of the binary over $\sim$18 yr, implying that the total masses are low.

We also find a point source which could be the embedded protostar's photosphere in the LBT AO [\ion{Fe}{2}] $1.644 \mu$m continuum. Assuming no extinction and converting $H$ photometry into stellar mass, we find that LV 1 may have one very-low-mass star with a brown dwarf secondary, or even be a double brown dwarf. 

The magnetopause of 168--326 NW is estimated to be 110 AU from the proplyd, which reasonably agrees with the location of the bow shock seen in H$\alpha$. The morphology of proplyds may help understand the magnetic field strength and profile.

\acknowledgements
We thank the referee for helpful comments that greatly improved this paper. We thank the whole Magellan Staff for making this wonderful telescope possible. We especially thank Povilas Palunas (for help over the entire MagAO commissioning run). Juan Gallardo, Patricio Jones, Emilio Cerda, Felipe Sanchez, Gabriel Martin, Maurico Navarrete, Jorge Bravo and the whole team of technical experts helped do many exacting tasks in a very professional manner. Glenn Eychaner, David Osip and Frank Perez all gave expert support which was fantastic. Also thanks to Victor, Maurico, and Hugo for running the telescope so well. It is a privilege to be able to commission an AO system on such a fine telescope and site. Thanks from the whole MagAO team. And thanks to Miguel Roth, Francisco Figueroa, Roberto Bermudez, Sergio Veliz, Mark Leroy and, of course, Mark Philips for making this all happen---and very smoothly---despite the large AO team that was needed at the mountain and all the headaches and extra work it created for the LCO staff. We also thank the teams at Steward Observatory Mirror Lab, Microgate and ADS for building such a great ASM. The MagAO system was developed with support from the NSF MRI and TSIP programs. The VisAO camera was developed with help from the NSF ATI program. Y.L.W.'s and L.M.C's research were supported by NSF AAG and NASA Origins of Solar Systems grants. J.R.M. is grateful for the generous support of the Phoenix ARCS Foundation. K.M. was supported under contract with the California Institute of Technology (Caltech) funded by NASA through the Sagan Fellowship Program. 

The LBT is an international collaboration among institutions in the United States, Italy and Germany. LBT Corporation partners are: the University of Arizona on behalf of the Arizona university system; Istituto Nazionale di Astrofisica, Italy; LBT Beteiligungsgesellschaft, Germany, representing the Max-Planck Society, the Astrophysical Institute Potsdam, and Heidelberg University; the Ohio State University, and the Research Corporation, on behalf of the University of Notre Dame, University of Minnesota and University of Virginia.

\clearpage
\begin{deluxetable}{lcc}
\tabletypesize{\scriptsize}
\tablecaption{Properties of LV 1 Binary\label{tbl-1}}
\tablewidth{0pt}
\tablehead{
\colhead{Property} &
\colhead{SE} &
\colhead{NW}
}
\startdata
Separation (\arcsec)\tablenotemark{a} & $\cdots$ & $0.396\pm0.003$\\
P.A. (\arcdeg)\tablenotemark{a} & $\cdots$ & $325.6\pm 0.3$\\
$H$ (mag)\tablenotemark{b} & $13.15\pm0.21$ & $13.33\pm0.21$\\
$M_H$ (mag)\tablenotemark{c} & $5.82\pm0.30$ & $6.00\pm0.30$\\ 
Mass($M_{\sun}$) & $0.073^{+0.012}_{-0.011}$ & $0.067^{+0.011}_{-0.009}$\\
\enddata
\tablenotetext{a}{$\mbox{LBT}$ AO 1.644$\mu$m on 2011 October 15 (UT) \citep{C12b}.}
\tablenotetext{b}{\cite{P98}.}
\tablenotetext{c}{$D = 414\pm7$ pc \citep{M07}.}
\end{deluxetable}

\clearpage

\end{document}